\documentclass[preprint,12pt,review]{elsarticle}

\usepackage{amssymb}
\usepackage{amsmath}

\usepackage{bm}
\usepackage{mathtools}
\usepackage{caption}
\usepackage{graphicx}
\usepackage{latexsym}
\usepackage{booktabs}
\usepackage{multirow}
\usepackage{algorithm}
\usepackage{algpseudocode}
\usepackage{physics}
\usepackage{xcolor}
\usepackage{pgf}
\usepackage{braket}
\usepackage{autobreak}
\usepackage{makecell}
\usepackage{verbatim}
\usepackage{geometry}
\usepackage{subcaption}
\usepackage{hyperref}\hypersetup{hidelinks}
\usepackage{cleveref} 
\usepackage{amssymb}
\graphicspath{{Fig/}}
\graphicspath{{Fig/template/}}

\newcommand{\lrr}[1]{\left(#1\right)}
\newcommand{\lrs}[1]{\left[#1\right]}
\newcommand{\lrc}[1]{\left\{#1\right\}}

\usepackage{lineno}

\journal{}

\begin{document}

\begin{frontmatter}



\title{Airfoil shape optimization via coherent Ising machine} 
 

\author[fir,sec]{Hao Ni} 
\author[fou]{Qi Gao}
\author[fir]{Zhen Lu\corref{cor1}}\ead{zhen.lu@pku.edu.cn}
\author[fir,thr]{Yue Yang\corref{cor1}}\ead{yyg@pku.edu.cn}
\cortext[cor1]{Corresponding author.}

\address[fir]{State Key Laboratory for Turbulence and Complex Systems, School of Mechanics and Engineering Science, Peking University, Beijing 100871, China}
\address[sec]{College of Engineering, Peking University, Beijing 100871, China}
\address[fou]{Beijing QBoson Quantum Technology Co., Ltd., Beijing 100015, China}
\address[thr]{HEDPS-CAPT, Peking University, Beijing 100871, China}

\begin{abstract}


Airfoil shape optimization presents a challenge where classical solvers frequently struggle with computational efficiency and local minima. 
In the promising paradigm of quantum computing, the coherent Ising machine (CIM), a specialized physical solver, offers acceleration capabilities. However, its native discrete binary architecture restricts the application in aerodynamic design. 
To bridge this gap, we propose a comprehensive framework that translates airfoil shape optimization into hardware-compliant quadratic unconstrained binary optimization formulations. 
We integrate high-order response surface models via the Rosenberg order reduction, enabling the CIM to capture strong nonlinearities in the aerodynamic performance response. 
Furthermore, we introduce a block-diagonal scalarization strategy that compose trade-off scenarios into a single optimization. 
Validated on the NACA 4-digit airfoil series using CIM hardware with 615 spins, the framework successfully locates the global optimum with a computational speedup of three orders of magnitude compared to the classical simulated annealing. 
The parallel embedding capacity allows for the extraction of an entire optimal Pareto front in a single hardware execution. 
This work demonstrates a viable, quantum-enhanced paradigm for engineering optimization. 

\end{abstract}

\begin{keyword}
Quantum computing \sep  Airfoil shape optimization \sep Coherent Ising machine \sep Quadratic unconstrained binary optimization \sep Response surface model



\end{keyword}

\end{frontmatter}



\newpage
\section{Introduction}
\label{sec:intro}

Optimization in computational mechanics is a foundational pillar of engineering design, driving advancements across structural~\cite{Liu2026,Li2026}, thermal~\cite{Jaluria2019}, and fluid systems~\cite{Mohammadi2004}.
Within fluid dynamics specifically, airfoil shape optimization aims to identify geometric profiles that maximize performance metrics, such as the lift-to-drag ratio, under operational constraints~\cite{Jameson1988Aerodynamic}. 
The fundamental challenge lies in the intricate nature of the design space. 
Parameterizing geometries yields a high-dimensional search space~\cite{Samareh2001Survey}, and fluid-structure interactions~\cite{Yang2023,Fan2026} introduce strong nonlinearities in the aerodynamic performance response. 
Furthermore, practical design often involve multi-objective trade-offs~\cite{Marler2004Survey}. 
Together, these factors create a highly non-convex objective landscape that is difficult to navigate~\cite{Forrester2009Recent}. 

Over the past few decades, classical numerical optimization methods~\cite{Skinner2018State} have been extensively applied to airfoil design. 
Gradient-based algorithms, typically coupled with computational fluid dynamics (CFD) solvers via adjoint methods, facilitate efficient local searching~\cite{Giles2000Introduction}. 
Although such methods converge rapidly for small-amplitude shape modifications, they are prone to fall into suboptimal local extrema~\cite{Giannakoglou2002} and sensitive to numerical noise~\cite{Martins2021Engineering}. 
In contrast, global optimization strategies, such as genetic algorithms~\cite{Obayashi1997Multiobjective} and simulated annealing (SA)~\cite{Kirkpatrick1983}, are widely adopted due to their gradient-free nature and ability to escape local minima~\cite{Giannakoglou2002}, but they rely on sequential, probabilistic sampling. 
Consequently, traversing an exponentially growing, high-dimensional parameter space requires a massive number of iterations. 
Even when the computational burden of CFD is bypassed using surrogate models~\cite{Forrester2009Recent}, the inherent time complexity of these classical algorithms remains a bottleneck, highlighting the need for faster optimization paradigms. 

Quantum computing has emerged as a revolutionary paradigm for aerospace engineering applications~\cite{Givi2020,Wang2023,Liu2024Towards}. 
By exploiting quantum phenomena such as superposition and entanglement, quantum processors can theoretically process a vast number of solutions, offering the potential for exponential speedup~\cite{Daley2022}.
Existing studies have explored quantum algorithms~\cite{Montanaro2016a,Lubasch2020,Jin2024,Xu2025,An2023,An2025}, ranging from CFD to structural topology optimization~\cite{Montanaro2016b,Wang2020,Todorova2020,Liu2023,Bharadwaj2023,Jaksch2023,Pfeffer2023,Schalkers2024,Chen2024,Sanavio2024,Lu2024Quantum,Meng2024,Wang2025Quantum,Tang2025,Zhang2025,Wawrzyniak2025,Xiao2025,Xu2024,Xu2024b,Xu2025Hybrid,Liu2024Quantum,Xu2026,Li2023}.
Although universal gate-based quantum computers offer immense theoretical power, current devices in the noisy intermediate-scale quantum era~\cite{Preskill2018} remain constrained by high error rates, decoherence, and limited qubit availability~\cite{Meng2025Challenges,Wright2024,Bharadwaj2025,Song2025,Wang2026,Yang2026}. 
Consequently, specialized physical solvers designed specifically to minimize quadratic unconstrained binary optimization (QUBO) or Ising Hamiltonian models have emerged as more accessible and immediately viable platforms for near-term engineering optimization~\cite{Johnson2011,King2023,McMahon2016,Inagaki2016,Wei2026Versatile}.

The coherent Ising machine (CIM) is a quantum-classical hybrid computing architecture to solve the QUBO efficiently~\cite{McMahon2016,Inagaki2016}. 
By leveraging networks of degenerate optical parametric oscillators (DOPOs)~\cite{Wang2013,Marandi2014}, CIMs physically simulate the Ising spin systems to rapidly locate optimal ground-state energy configurations. 
A fundamental advantage of the CIM lies in its specialized measurement-feedback control system, which enables fully programmable, all-to-all connections among thousands of optical spins~\cite{McMahon2016}. 
This natively dense connectivity allows the optical network to encode highly coupled, dense graph structures without suffering from the resource-intensive minor-embedding overheads required by the sparse, localized topologies of superconducting quantum annealers~\cite{Hamerly2019}. 
Consequently, for large-scale combinatorial optimization challenges, empirical benchmarks demonstrate that CIMs exhibit superior time-to-solution scaling for dense problems compared to traditional quantum annealers~\cite{Hamerly2019}, while accelerating computation by several orders of magnitude over classical SA~\cite{Honjo2021}.

Driven by these hardware advantages, CIMs and broader QUBO formulations have been deployed across various tasks. 
Applications range from foundational graph problems~\cite{Hamerly2019,Honjo2021,Takesue2025} to practical scientific and engineering challenges~\cite{Zha2023Encoding,Wen2026Largescale,Ye2023Quantum,Raisuddin2022,Nguyen2024,Asztalos2024Reducedorder,Kuya2024Quantum,Asaga2025Obtaining,Wang2024,Yuan2023,Sukulthanasorn2025Novel,Kitai2020Designing,Suzuki2026,Kuya2025Quantum,Tamura2025Blackbox}, including molecular docking~\cite{Zha2023Encoding,Wen2026Largescale}, finite element simulations~\cite{Raisuddin2022,Nguyen2024}, structural topology and truss optimization~\cite{Sukulthanasorn2025Novel,Suzuki2026}, and artificial intelligence~\cite{li2025unified,li2025quantum,li2026exact}. 
However, adapting natively binary Ising models to the continuous, high-dimensional domain of aerodynamic design remains profoundly challenging.
Recent efforts have explored airfoil shape optimization using Ising hardware coupled with simple quadratic surrogate models~\cite{Kuya2025Quantum,Tamura2025Blackbox}. 
However, forcing complex fluid dynamics problems into quadratic forms oversimplified the problem, limiting the capture of strong nonlinearities. 
Furthermore, these preliminary demonstrations were restricted to simple, single-objective test cases.  
Realistic aerodynamic design, however, is inherently driven by competitive, multi-objective trade-offs. 
Therefore, establishing a comprehensive optimization framework to map continuous, high-order, and multi-objective aerodynamic landscapes into hardware-compliant discrete architectures remains a critical challenge.

To address these limitations, we propose a comprehensive, closed-loop framework for airfoil shape optimization using the CIM. 
We map the aerodynamic metrics into discrete binary formulations using polynomial response surface models (RSMs). 
Our framework incorporates a Rosenberg order reduction, enabling the embedding of fourth-order RSMs to capture strong physical nonlinearities. 
Furthermore, to navigate the multi-objective trade-offs, we introduce a parallel embedding architecture. 
This strategy maps multiple design preference scenarios simultaneously, allowing the extraction of an optimal Pareto front in a single hardware execution. 
Validated on the NACA 4-digit airfoil series using the largest commercial-grade CIM hardware to date, this approach demonstrates orders-of-magnitude computational speedups over classical heuristic algorithms while preserving physical fidelity. 
To the best of our knowledge, this research represents the first fully integrated application of the CIM hardware to continuous design in computational mechanics, establishing a complete, end-to-end pipeline from high-order aerodynamic modeling to physical hardware execution.

The remainder of this paper is organized as follows. 
Section~\ref{sec:method} details the formulations of our end-to-end optimization framework using the CIM.
Section~\ref{sec:result} presents the experimental validations and hardware benchmarking on the NACA 4-digit airfoil series.  
Section~\ref{sec:conclusion} provides the concluding remarks and discusses future hardware scaling.

\section{Methods}
\label{sec:method}

\begin{figure}
    \centering
    \includegraphics[width=\linewidth]{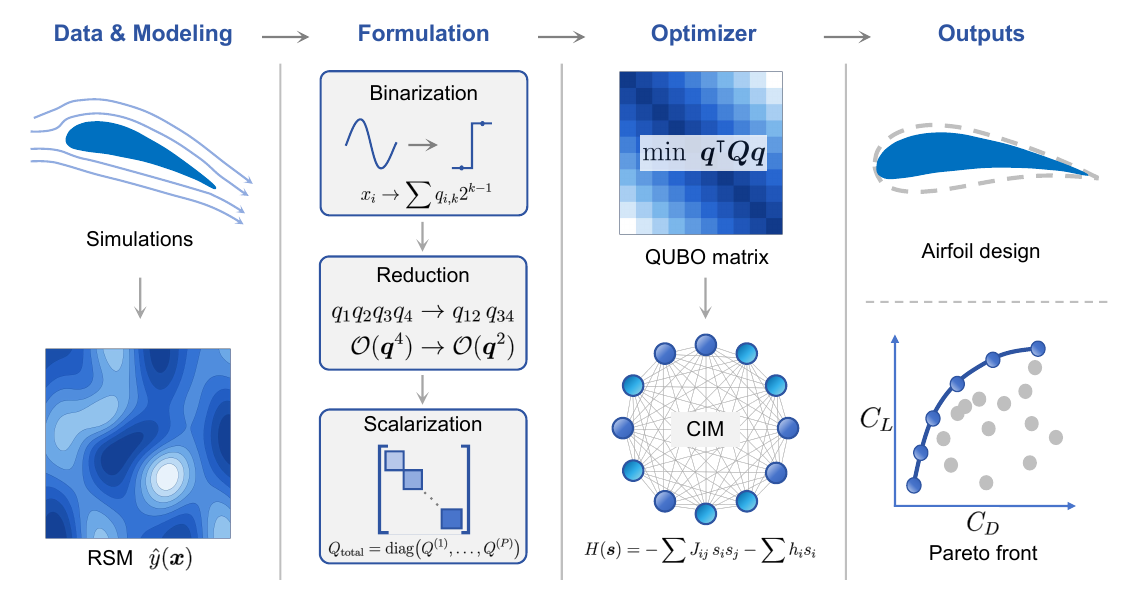}
    \caption{End-to-end airfoil shape optimization workflow using CIM. The process bridges classical aerodynamic simulation with quantum-enhanced optimization, progressing from data generation and surrogate modeling to binary formulation, hardware embedding, and final design decoding.
    }
    \label{fig:workflow}
\end{figure}

We present a comprehensive framework for airfoil shape optimization using CIM. 
As illustrated in Fig.~\ref{fig:workflow}, our workflow has four stages. 
First, classical simulations generate a training dataset, which is used to construct polynomial RSMs of the aerodynamic performance metrics. 
Second, these continuous surrogates are transformed into discrete binary optimization problems. 
Third, the optimization problems are solved on the CIM hardware, where the optical network evolves to identify low-energy spin configurations. 
Finally, the binary solutions are mapped back to physical airfoil geometries, or reconstructed into the Pareto front for multi-objective trade-off analysis.

\subsection{CIM-based optimization}
\label{sec:CIM}

The CIM is a network of DOPOs that physically simulates the Ising spin model~\cite{Wang2013,Marandi2014,McMahon2016}. 
Unlike gate-based quantum computers, the CIM evolves the network of optical pulses to find the ground state of a user-defined Hamiltonian. 
The optimization problem is formulated as finding the spin configuration $\bm{s}\in\lrc{-1,1}^N$ that minimizes the Ising Hamiltonian
\begin{equation}\label{eq:Ising}
H(\boldsymbol{s}) = -\sum^N_{i < j} J_{ij} s_i s_j - \sum^N_{i=1} h_i s_i, 
\end{equation}
where $N$ denotes the number of spins, $h_i$ represents the local bias fields, and $J_{ij}$ denotes the coupling strengths between spins. 
The measurement-feedback system of CIM allows for dense all-to-all connectivity, enabling the embedding of complex optimization graphs~\cite{McMahon2016}. 

To apply the CIM to engineering design, the problems are mapped into the binary domain. 
This yields the QUBO model 
\begin{equation}\label{eq:QUBO}
    H(\bm{q}) = \sum^N_{i \le j} Q_{ij} q_i q_j
\end{equation}
in terms of binary variables $\bm{q}\in\lrc{0,1}^N$. 
It is equivalent to the Ising model in Eq.~\eqref{eq:Ising} via the transformation $s_i=2q_i-1$.
From a hardware perspective, both the physical spins $s_i$ and the logical binary variables $q_i$ correspond to the fundamental computing nodes of the system, serving as the equivalent of qubits in quantum computing paradigms.
Consequently, any problem formulated as a QUBO can be directly solved using the CIM hardware.
Furthermore, compared to other quantum computing paradigms, the CIM offers unique advantages including large qubit scale and room-temperature operation~\cite{Honjo2021,Wei2026Versatile}, which facilitates solving specific engineering problems. 

\subsection{Airfoil design and surrogate modeling}
\label{sec:airfoil}

The primary objective of this study is to optimize the airfoil geometry, parameterized by a set of design variables $\bm{x}$, such as the location, maximum camber, and maximum thickness. 
The optimization problem seeks to identify the optimal configuration $\bm{x}^*$ that maximizes the objective function $y\lrr{\bm{x}}$ on aerodynamic performance, such as the lift-to-drag ratio. 
Direct evaluation of $y\lrr{\bm{x}}$ via CFD is computationally expensive~\cite{Forrester2009Recent} and prone to numerical noise~\cite{Giannakoglou2002}. 
Thus, we employ polynomial RSMs~\cite{Myers2016Response} as surrogates $\hat{y}\lrr{\bm{x}}$. 
The construction of surrogates involves a trade-off between modeling fidelity and computational resource consumption on quantum hardware.

We first employ a second-order RSM
\begin{equation}\label{eq:RSM2}
\hat{y}^{(2)}(\boldsymbol{x}) = \beta_0 + \sum_{i=1}^{n} \beta_i x_i + \sum_{i=1}^{n} \sum_{i \le j}^{n} \beta_{ij} x_i x_j,
\end{equation}
where $n$ denotes the number of design variables and $\bm{\beta}$ are coefficients obtained via regression. 
This formulation captures the primary curvature of the design space and maps to the QUBO format in Eq.~\eqref{eq:QUBO}. 
Due to its simplicity, the discrete optimization of this model requires minimal auxiliary spins, facilitating the efficient embedding of large-scale or multi-objective problems.

Since aerodynamic phenomena often introduce strong non-linearities that second-order models fail to capture accurately, we extend the framework to include a fourth-order RSM
\begin{equation}\label{eq:RSM4}
\hat{y}^{(4)}(\boldsymbol{x}) = \hat{y}^{(2)}(\boldsymbol{x}) + \sum_{i,j,k} \beta_{ijk} x_ix_jx_k + \sum_{i,j,k,l} \beta_{ijkl} x_ix_jx_kx_l.
\end{equation}
Although this higher-order model improves the goodness-of-fit, it introduces higher-order interaction terms. 
Mapping these terms to the pairwise Ising architecture requires order reduction techniques, which inevitably consume additional spins. 
Therefore, the selection of the RSM order is determined by balancing the requirement for modeling accuracy against the available spin capacity of the hardware.

\subsection{Binary encoding and order reduction}
\label{sec:encode}

We transform the continuous design variables $\bm{x}$ to the binary variables of QUBO in Eq.~\eqref{eq:QUBO} using a fixed-point encoding scheme. 
Each design variable $x_i \in [L_i, U_i],~i=1,\cdots,n$, with lower and upper bounds $L_i$ and $U_i$, is discretized using $K_i$ binary variables. 
The mapping from the binary vector $\bm{q}_i$ to $x_i$ is
\begin{equation}\label{eq:encoding}
x_i = L_i + \frac{U_i - L_i}{2^{K_i}-1} \sum_{k=1}^{K_i} q_{i,k} 2^{k-1}, \quad q_{i,k} \in \{0,1\}.
\end{equation}
Substituting Eq.~\eqref{eq:encoding} into the RSMs transforms the continuous optimization problem into a Boolean optimization. 
When applied to the second-order RSM in Eq.~\eqref{eq:RSM2}, the resulting polynomial contains terms up to degree two in $\bm{q}$, allowing direct compilation onto the CIM using $N=\sum^n_{i=1} K_i$ spins.

However, applying the same encoding to the fourth-order RSM in Eq.~\eqref{eq:RSM4} introduces high-order interaction terms up to degree four, e.g., $q_1q_2q_3q_4$, which are not natively supported by the CIM. 
We employ the Rosenberg reduction method~\cite{Rosenberg1975} to transform these high-order terms into quadratic forms, by introducing auxiliary binary variables, e.g., $q_{12}=q_1q_2$, and enforcing consistency via the penalty terms $ H_p = \lambda\lrr{q_1q_2-2q_1q_{12}-2q_2q_{12}+3q_{12}}$ in the Hamiltonian. 
Here, $\lambda$ is a penalty coefficient sufficiently large to ensure the constraints are satisfied, and it should be finite, as an excessively dominant penalty term can lead to numerical instability and diminish the precision of the objective function's energy landscape~\cite{Glover2022Quantum}. The specific criteria and derivation for selecting an optimal $\lambda$ are detailed in~\ref{sec:penalty}. 
This order reduction converts a polynomial RSM of arbitrary order into a QUBO form. 

Note that explicitly mapping these high-order interactions incurs a significant qubit overhead. 
Crucially, every reduction step consumes physical spins to represent the auxiliary variables. 
Consequently, the total number of spins required scales not only with the encoding precision $K$ but also heavily with the density of higher-order terms in the RSM. 
This creates a hardware trade-off where pursuing higher modeling fidelity or discretization resolution rapidly consumes the limited spin budget.

While order reduction increases spin consumption, the resulting dense QUBO graph inherently favor the CIM's architecture. Sparse quantum annealers suffer from an $O(N^2)$ embedding qubit overhead and severe time-to-solution degradation on such dense graphs~\cite{Hamerly2019}. In contrast, the CIM provides intrinsic all-to-all connectivity eliminating embedding penalties. Consequently, the CIM achieves an empirical computation time complexity of roughly $O(N)$ to reach a given solution quality in representative problems, outperforming classical SA, which scales as $O(N^2)$~\cite{Honjo2021}.

\subsection{Multi-objective optimization}
\label{sec:multi}

Aerodynamic design typically involves trade-offs between different performance metrics, which can be formulated as a multi-objective optimization problem. 
The goal is to simultaneously optimize a vector of $M$ objective functions $\lrs{y_1(\boldsymbol{x}), y_2(\boldsymbol{x}),\dots, y_M(\boldsymbol{x})}^\intercal$. 
To assess trade-offs, it is essential to approximate the Pareto front, the set of optimal compromises where improving one objective is impossible without sacrificing other objectives~\cite{Miettinen1999Nonlinear,Nemec2004Multipoint}. 

To enable compatibility with the scalar Ising Hamiltonian of the CIM, we employ a scalarization strategy. 
The problem is transformed into a single function $y\lrr{\bm{x};\bm{w}} = \sum_{i=1}^{M} w_i y_i(\bm{x})$ via a normalized weight vector $\bm{w}=\lrs{w_1,\cdots,w_M}^\intercal$, with $\sum_{i=1}^{M} w_i = 1$. 
This scalarized objective $y\lrr{\bm{x};\bm{w}}$ is then mapped to a QUBO problem using the encoding scheme described in Sec.~\ref{sec:encode}. 
The aggregated QUBO matrix is equivalent to the linear combination
\begin{equation}\label{eq:QUBOM}
    \bm{Q}\lrr{\bm{w}} = \sum_{i=1}^{M} w_i \bm{Q}_i,
\end{equation}
where $\bm{Q}_i$ denotes the QUBO matrices derived from each individual objective with the aforementioned methodology. 
The detailed derivation is provided in~\ref{sec:scalarization}.

Tracing the Pareto front requires solving the QUBO problem in Eq.~\eqref{eq:QUBOM} for a sequence of weight vectors, representing varying trade-offs between the objectives.
On classical solvers, the computational time scales linearly with the number of weight vectors.  
To efficiently explore the Pareto front, we propose a parallel embedding architecture, instead of solving the QUBO problems for each weight vector. 
We construct a block-diagonal composite QUBO matrix
\begin{equation}\label{eq:Qt}
\boldsymbol{Q}_{\text{total}} = \text{diag}\left[ \boldsymbol{Q}(\boldsymbol{w}^{(1)}), \boldsymbol{Q}(\boldsymbol{w}^{(2)}), \ldots, \boldsymbol{Q}(\boldsymbol{w}^{(P)}) \right]
\end{equation}
for set of weight vectors $\{\boldsymbol{w}^{(i)}\}_{i=1}^{P}$. 
It embeds $P$ distinct scenarios onto the CIM hardware simultaneously. 
By utilizing the optical network's bandwidth, we achieve $O(1)$ time complexity relative to the number of Pareto points sampled. 
This allows the entire trade-off curve to be generated in few hardware runs.

\section{Results}
\label{sec:result}

We validate the proposed optimization framework using three distinct case studies. 
The computational complexity and hardware resource demands of these problems are primarily determined by the dimensionality of the discrete parameter space, the order of RSM, and the number of optimization objectives. 
Although the method in Sec.~\ref{sec:method} supports scalability across all these aspects, current physical limitation of the CIM hardware regarding spin count make it infeasible to fully incorporate all advantages into a single validation case.

Therefore, each case focuses on one specific aspect to characterize the method's capability. 
Section~\ref{sec:case_QUBO} validates the efficiency by varying over all design variable of the NACA 4-digit airfoil family to identify the optimal design from $2^{24}\approx 1.68\times 10^7$ candidates using a second-order RSM. 
Section~\ref{sec:case_HOBO} addresses the representation of strong aerodynamic nonlinearities by employing a fourth-order RSM to obtain solutions that are more accurate than those approximated from a second-order RSM.
Section~\ref{sec:case_multi} extends to multi-objective optimization to illustrate the framework's advantage in tracing the Pareto front. 
This modular validation confirms that with continued advances in CIM hardware, our framework is able to solve problems combining high variable dimensions, high-order RSM, and multiple objectives. 

\subsection{Problem setup}
\label{sec:setup}

To evaluate the aerodynamic performance, we employed XFOIL v6.99~\cite{Drela1989}, a tool widely adopted for analysis of airfoils. 
All simulations were conducted under representative low-speed incompressible conditions with a Reynolds number of $Re=3.0\times 10^6$ and a fixed angle of attack $\alpha = 5.0^\circ$. 
The maximum iteration limit for each XFOIL computation was set to 3000 to achieve numerical convergence and high-fidelity data.
To ensure geometric feasibility during the random sampling phase, design variables were constrained within the NACA 4-digit parameters, with the maximum camber $A\in\lrs{0,6}\%$, the location of maximum camber  $B\in\lrs{2,5}\%$, and the maximum thickness $T\in\lrs{6,20}\%$ of the chord length $c$. 
This bounded cubic design space prevents the generation of invalid geometries while providing a sufficiently complex landscape for the optimization framework.

We executed the quantum-enhanced optimizations on the QBoson CIM~\cite{Wei2026Versatile}, a photonic computing platform supporting up to 1000 spins. 
Hardware interaction was managed via the Kaiwu SDK v1.3.0~\cite{KaiwuSDK}. 
To evaluate CIM performance, we implemented three classical methods for comparison. 
First, a brute-force (BF) exhaustive search was performed to establish the ground truth global optimum, enabling exact verification of solution accuracy. 
Second, the gradient descent (GD) was employed as a classical local search strategy. 
Finally, the SA was utilized as the classical counterpart to CIM. 
As a heuristic approach that also solves QUBO problems, SA serves as the primary benchmark for evaluating the CIM's relative convergence efficiency and solution quality.
We empirically set SA hyperparameters as initial temperature of 5000, minimum temperature of 0.001, cooling rate of 0.9, and 50 iterations per temperature level, to balance solution accuracy and computational efficiency.
All the classical methods (BF, GD, and SA) were executed on a classical desktop computer with a 1.20 GHz CPU and 32 GB memory.

\subsection{Baseline validation}
\label{sec:case_QUBO}

\begin{figure}[ht]
    \centering
    \includegraphics[width=\linewidth]{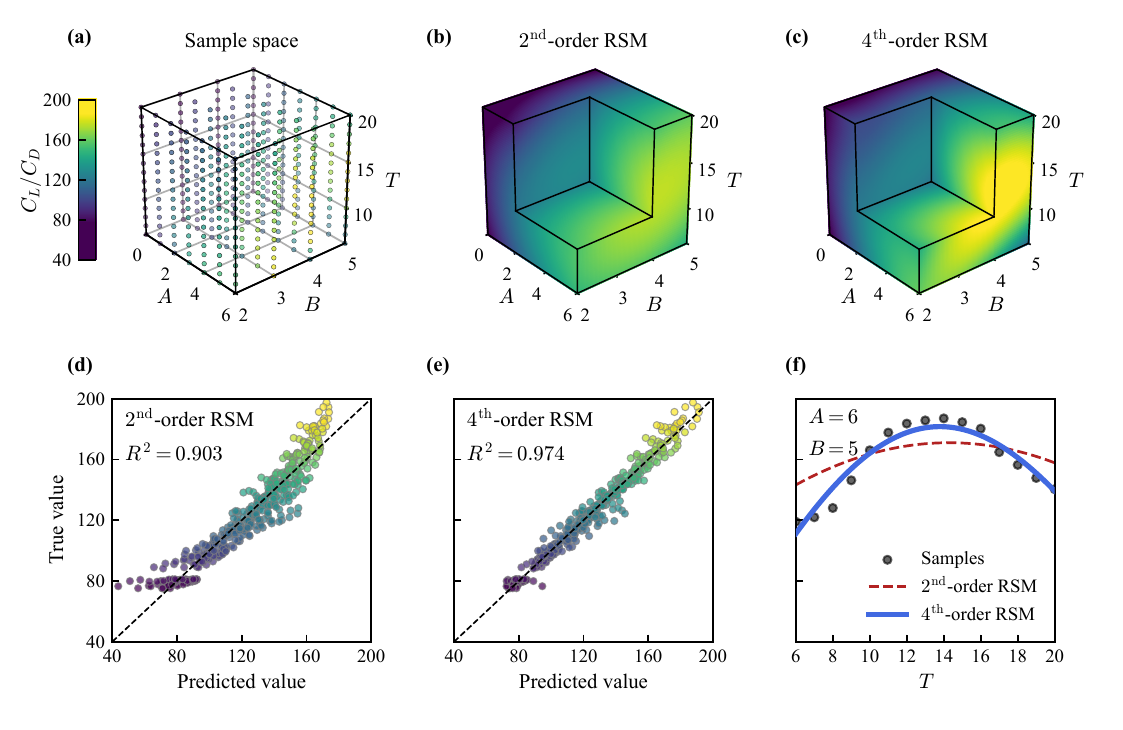}
    \caption{
    Sampling, construction, and validation of RSMs for the NACA 4-digit airfoil series.
    (a)~3D visualization of the design space and sampling points, colored by the lift-to-drag ratio $C_L/C_D$.
    (b, c)~Volumetric representation of the second- and fourth-order RSMs displayed with a partial cutaway for interior visualization.
    (d, e)~Parity plots of predicted vs. true values of $C_L/C_D$. 
    (f)~2D cross-section at fixed camber parameters $A=6$ and $B=5$, showing the dependence of $C_L/C_D$ on thickness $T$.
    }
    \label{fig:rsm}
\end{figure}

This case study evaluates the framework's capability to navigate a high-dimensional discrete search space. 
The optimization objective is to maximize the lift-to-drag ratio $C_L/C_D$ of the airfoil. 
To construct the RSM, we selected all integer grid points within the parameter space, $A\in\lrc{0,\cdots,6}$, $B\in\lrc{2,\cdots,5}$, and $T\in\lrc{6,\cdots,20}$, as sampling points. 
We then utilized XFOIL to calculate  $C_L/C_D$ for each sample point, as plotted in Fig.~\ref{fig:rsm}a. 
These data points were used to regress both second- and fourth-order RSMs, as shown in Figs.~\ref{fig:rsm}b and \ref{fig:rsm}c, respectively. 
The parity plots in Figs.~\ref{fig:rsm}d and \ref{fig:rsm}e show that the second-order RSM captures the global trend, yielding a goodness of fit $R^2=0.903$,  but exhibits variance at both extremes of the $C_L/C_D$ regions. 
Meanwhile, the fourth-order RSM provides a better fit with $R^2=0.974$. 
Figure~\ref{fig:rsm}f further highlights this behavior in a 2D cross-section at $A=6$ and $B=5$, selected to evaluate model performance near a performance peak. Here, samples refers to the XFOIL-evaluated data subset within this cross-section. We observe that the second-order RSM smooths out the peak, whereas the fourth-order model accurately tracks the variance of $C_L/C_D$.

For this baseline efficiency test, we employ the second-order RSM. 
Although less accurate than the fourth-order RSM, its native compatibility with CIM makes it the suitable for benchmarking without the overhead of order-reduction auxillary spins. 
We discretized the design variables using $K=8$-bit binary encoding. 
This generates a discrete search space containing $2^{24}$ candidate designs. 
To map this problem onto the CIM, we employed the precision adaptive split (PAS) to handle the integer-resolution (see~\ref{sec:split}).
This process introduced auxiliary qubits to split large matrix elements while preserving implementation accuracy, resulting in a final 615-spin Ising Hamiltonian. 

\begin{figure}[!ht]
    \centering
    \includegraphics[width=\linewidth]{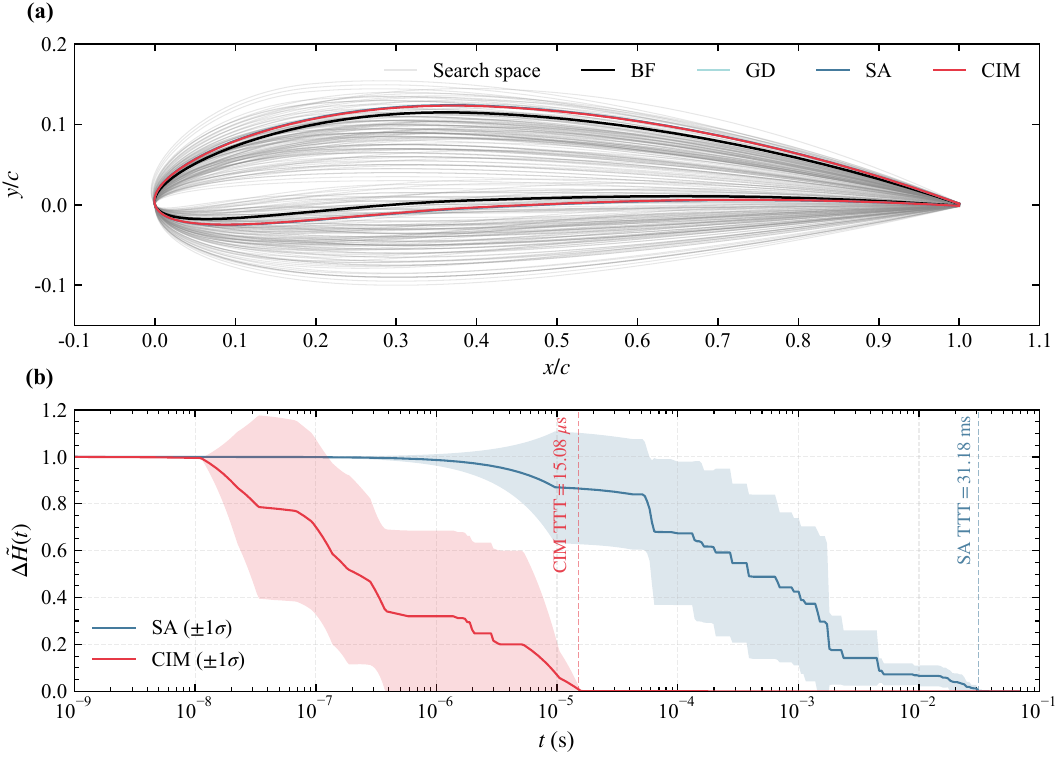}
    \caption{Validation of optimization fidelity and computational efficiency on the baseline case with $Re=3.0\times 10^6$ and $\alpha = 5.0^\circ$. 
    (a)~Geometric profiles of the optimized airfoils identified by the CIM (red), overlaid with results from SA (light blue), GD (dark blue), and the ground truth BF (black). 
    Thin gray lines represent a subset of the airfoil sample space. 
    (b)~Temporal convergence comparing between the CIM and SA. 
    The optimization trajectories of the CIM (red line) and the SA baseline (blue line) is tracked via the normalized Hamiltonian gap $\Delta \tilde{H}(t)$. 
    The shaded regions denote the standard deviation across five independent runs, and vertical dashed lines mark the TTTs.}
    \label{fig:baseline}
\end{figure}

Figure~\ref{fig:baseline} compares the optimization results.
The CIM demonstrates high solver fidelity. 
As shown in Tab.~\ref{tab:baseline}, the optimal design variables identified by the CIM are nearly identical to those found by GD and SA. 
The airfoil profiles obtained from different methods inherently overlap in Fig.~\ref{fig:baseline}a. 
This confirms that the CIM successfully finds the global optimum, performing at the same level with classical solvers. 
On the other hand, a notable deviation exists between the optimizers (GD, SA, and CIM) and the ground truth BF solution. 
Since the BF method exhaustively searches the underlying physics space, this discrepancy highlights the limitation of the expressibility of the second-order RSM. 

\begin{table}[htbp]
    \centering
    \caption{Optimal design parameters obtained from different solvers.}
    \label{tab:baseline}
    \begin{tabular}{l c c c c}
        \toprule
          & BF & GD & SA & CIM \\
        \midrule
        $A$ (\%$c$) & 6.00 & 6.00 & 6.00 & 6.00 \\
        $B$ (\%$c$) & 3.92 & 4.15 & 4.14 & 4.15 \\
        $T$ (\%$c$) & 11.22 & 13.13 & 13.14 & 13.03 \\
        \bottomrule
    \end{tabular}
\end{table}

We benchmarked the computational efficiency of the CIM against the classical SA baseline by analyzing the time required to reach an equivalent solution quality. 
The convergence dynamics are quantified by the normalized Hamiltonian gap,
\begin{equation}\label{eq:DH}
    \Delta \tilde{H}(t) = \frac{H(t) - H_{\min}}{H_{\text{init}} - H_{\min}},
\end{equation}
where $H_{\min}$ is the theoretical minimum of the energy landscape and $H_{\text{init}}$ represents the initial Hamiltonian value.
Equation~\eqref{eq:DH} tracks the residual energy relative to the initial state. 
We define the time-to-target (TTT) as the physical wall-clock time required to reach a residual gap of $\Delta \tilde{H}(t) \le 1\%$, indicating convergence to the near-optimal region.
Figure~\ref{fig:baseline}b presents the average optimization trajectories over five independent runs. 

The CIM demonstrates a computational speedup of $2068$ times. 
Specifically, the CIM achieves the target threshold in approximately 15.08~$\mu$s, whereas the classical SA solver yields a TTT about 31.18~ms. 
Beyond the raw speedup, the shape of the convergence curves highlights the fundamental algorithmic differences. 
The SA exhibits a gradual, stepwise relaxation, characteristic of sequential Markov chain Monte Carlo processes that rely on thermal fluctuations to escape local minima~\cite{Kirkpatrick1983}.
In contrast, the CIM trajectory displays a precipitous drop in energy within the first few microseconds, intrinsic to the physics of the CIM. 
As the pump power increases, the system undergoes a pitchfork bifurcation where the mode configuration corresponding to the ground state experiences the maximum net gain, leading to a rapid, collective collapse into the low-energy solution~\cite{Leleu2019Destabilization}.

\subsection{Optimization using fourth-order RSM}
\label{sec:case_HOBO}

Although the baseline validation demonstrate the CIM's optimization capability, the second-order RSM fails to capture the strong nonlinearity in aerodynamic performance. 
To enhance physical fidelity, the optimization framework should account for high-order terms of the design parameters. 
In this case study, we optimize the maximum thickness $T$ of the NACA 4-digit airfoil series for $C_L/C_D$, with fixed camber parameters $A=6$ and $B=4$. 

\begin{figure}[ht]
    \centering
    \includegraphics[width=90 mm]{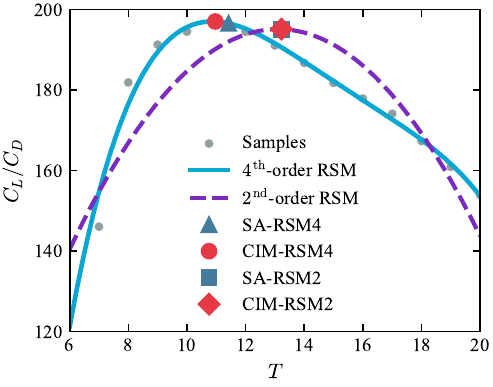}
    \caption{
    Optimization results for $C_L/C_D$ of the NACA 4-digit airfoils with $A = 6$ and $B = 4$ utilizing second- and fourth-order RSMs. 
    }
    \label{fig:hobo}
\end{figure}

Figure~\ref{fig:hobo} presents the difference in modeling fidelity between the second- and fourth-order RSMs. 
The response of sampled data exhibits a skewed distribution that cannot be adequately captured by a quadratic function. 
Consequently, the second-order RSM (dashed line) yields a poor fit of $R^2 = 0.79$, incorrectly shifting the predicted optimum to a higher $T$. 
In contrast, the fourth-order RSM (solid line) accurately reconstructs the varying curvature of the design space, achieving $R^2 = 0.98$. 
This comparison underscores the necessity of employing higher-order polynomial expansions for realistic aerodynamic design. 

We employed the Rosenberg reduction method~\cite{Rosenberg1975} to map the fourth-order RSM onto the CIM using a $K=5$-bit binary encoding.
This process introduces auxiliary logical spins to represent the higher-order interaction terms, resulting in a total problem size of 35 spins.
Effects of $K$ on accuracy and spin count are detailed in \ref{sec:split}. 
As shown in the optimization results in Fig.~\ref{fig:hobo}, the solution identified by the CIM using the fourth-order model (CIM-RSM4) aligns precisely with the true aerodynamic optimum observed in the sampling data. 
Conversely, while the CIM also found the mathematical maximum of the second-order model (CIM-RSM2), that solution deviates from the ground truth due to the model's intrinsic error. 
This result confirms that our framework effectively translates high-order aerodynamic performance nonlinearities into hardware-solvable forms. 

\subsection{Multi-objective optimization}
\label{sec:case_multi}

We consider the simultaneous maximizing $C_L$ and minimizing $C_D$. 
The problem is formulated using the scalarization strategy detailed in Sec.~\ref{sec:multi}, resulting the composite objective function
\begin{equation}\label{eq:CLCD}
    \max ~\frac{1}{1+w} C_L - \frac{w}{1+w} C_D,
\end{equation}
where $w$ is a weighting hyperparameter governing the preference between $C_L$ and $C_D$. 
To construct a Pareto front, we selected nine weight values $w \in \lrc{0, 5, 20, 50, 80, 100, 200, 500, 2000}$. 
The design space was defined by the NACA 4-digit airfoil series with fixed camber parameters $A = 6$ and $B = 4$, and variable $T \in [10, 15]$ sampled at an interval of 0.2. 
Utilizing a $K=$ 8-bit binary encoding, the problem was mapped to the CIM hardware. 

\begin{figure}[!ht]
    \centering
    \includegraphics[width=\linewidth]{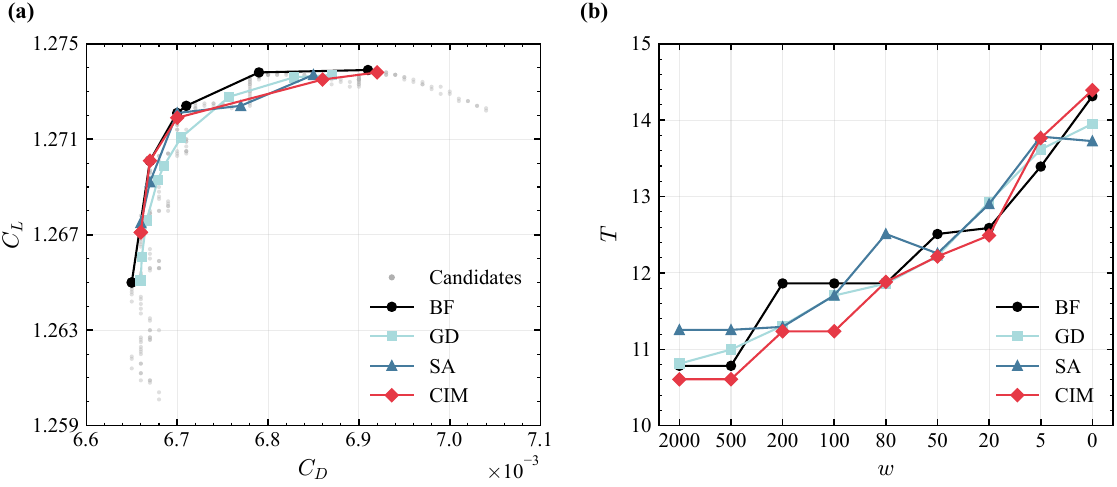}
    \caption{Multi-objective optimization for simultaneous maximizing $C_L$ and minimizing $C_D$. (a) Pareto fronts obtained by the BF, GD, SA, and CIM. Light gray dots represent the 256 candidate solutions in the discrete design space. Solid markers denote identified non-dominated Pareto optimal solutions. (b) Optimal thickness $T$ as a function of weight $w$. }
    \label{fig:pareto}
\end{figure}

We exploited the hardware's capacity to perform parallel one-shot optimization. 
A block-diagonal Hamiltonian containing the independent sub-problems for all nine weight vectors was constructed to enable simultaneous optimization, using 778 spins. 
This allowed to obtain all trade-off solutions in a single run, rather than executing the CIM nine separate times. 
Figure~\ref{fig:pareto} compares the Pareto fronts obtained by the CIM against the BF, GD, and SA. 
The CIM-derived front in Fig.~\ref{fig:pareto}a closely tracks the ground truth identified by the BF. 
This confirms that the parallel embedding strategy does not degrade solution quality. 

To verify the physical validity of these solutions, Fig.~\ref{fig:pareto}b plots the optimized $T$ against $w$. 
A clear monotonic trend is observed. 
As the weighting $w$ increase to prioritize drag reduction, the optimal $T$ decreases. 
Conversely, prioritizing lift with a higher $w$ drives the optimizer towards thicker airfoils. 
This behavior is aerodynamically consistent and optimization results successfully recovers this physical trade-off. 

\section{Conclusions} 
\label{sec:conclusion}

We presented a comprehensive framework for aerodynamic shape optimization utilizing the CIM. 
This approach exploits the hardware's advantages on solving the QUBO problems by transforming the continuous design variables into discrete binary formulations. 
The framework employs the Rosenberg order reduction to incorporate high-order polynomial RSMs, enabling modeling the strong nonlinearities in the objective landscape. 
Furthermore, we developed a matrix-level weighted scalarization strategy for multi-objective problems. 
It combines competing objectives within a unified formulation and enables a single hardware execution to generate a set of Pareto-front candidates. 
To the best of our knowledge, this constitutes the first closed-loop demonstration of the CIM hardware for engineering optimization tasks in computational mechanics, covering the full pipeline from modeling and reduction to hardware solving and solution decoding.

We validated the proposed framework through three case studies to assess optimization efficiency, model fidelity, and multi-objective capability. 
In the baseline optimization of a high-dimensional design space with $2^{24}$ candidates with 615 spins, the CIM demonstrated superior convergence characteristics compared to classical SA, locating the global optimum with a speedup of approximately three orders of magnitude. 
The optimal solutions obtained by the CIM were consistent with ground truth and classical benchmarks, confirming the solution fidelity. 
Furthermore, addressing the strong nonlinearities inherent in aerodynamic performances required extending the framework beyond the second-order model. 
By mapping the fourth-order RSM into QUBO, we successfully captured complex variations in lift-to-drag ratios that the second-order model failed to resolve.
This implies that although higher-order surrogates incur a spin overhead, they are essential for achieving the physical fidelity requisite for realistic engineering design.

The capability of the CIM was further demonstrated in multi-objective optimization. 
By employing a parallel embedding strategy for the scalarized objective functions, we retrieved the Pareto front in a single hardware execution. 
This ``one-shot'' approach contrasts with the classical sequential method, as it allows for the simultaneous exploration of multiple trade-off scenarios between lift and drag without iterative reprocessing. 
The resulting Pareto front aligned closely with the true aerodynamic trade-offs obtained by exhaustive search, validating that the superposition of scalarized weights does not degrade solution quality on the CIM.

By translating continuous field problems into discrete energy minimization tasks, this approach effectively bypasses the local minima traps inherent in gradient-based methods and the slow convergence of classical heuristic algorithms. 
However, the current study also highlights the physical constraints of contemporary CIM hardware. 
Due to the limited availability of spins, we validated efficiency, high-order modeling fidelity, and multi-objective trade-offs as separate case studies. 
A fully integrated optimization requires further extension of the spin capacity. 
Consequently, the practical advantage for extremely complex, coupled engineering systems remains to be verified on next-generation large-scale devices~\cite{Honjo2021}.

Looking forward, this framework offers a generalized pathway for solving complex optimization problems across computational mechanics. 
As hardware capacity scales, the methodology can be extended beyond 2D airfoils to high-dimensional problems such as 3D wing configuration, structural topology optimization, and material design~\cite{Sukulthanasorn2025Novel,Kitai2020Designing}.
This approach provides a new class of quantum-enhanced algorithms capable of solving nonlinear, non-convex engineering problems that are currently intractable for classical heuristics.

\section*{Acknowledgments}

We acknowledge the use of the QBoson cloud quantum computation platform for this work. 
We would like to thank Wenxin Li at Beijing QBoson Quantum Technology Co., Ltd. for discussions on CIM. 
This work has been supported in part by the National Natural Science Foundation of China (Nos.~12525201, 52306126, 12432010, and 12588201), the Beijing Natural Science Foundation (Nos.~QY25085 and F261001), and the National Key R\&D Program of China (No.~2023YFB4502600).

\appendix

\section{Penalty coefficient $\lambda$}
\label{sec:penalty}

The penalty coefficient $\lambda$ in the Rosenberg reduction must balance consistency and precision. 
An insufficient value permits constraint violation, e.g., $q_{ij} \neq q_i q_j$, yielding an infeasible QUBO formulation. 
Whereas an excessively large value compresses the energy resolution of the objective function, degrading optimization performance. 
To satisfy the constraints without over-penalization, $\lambda$ is set to just exceed the maximum objective-function energy change induced by any single-spin flip
\begin{equation}\label{eq:lambda}
    \lambda = \eta \max_{k \in \{1,\dots,N\}} \left| \Delta_k H(\boldsymbol{q}) \right|,
\end{equation}
where $\Delta_k H(\boldsymbol{q})$ denotes the energy change caused by flipping the $k$-th spin in the unpenalized objective and $\eta > 1$ is a safety margin. 
Because the minimum nonzero energy variation of the Rosenberg penalty term under a single-spin flip is exactly unity, this choice guarantees that the penalty always dominates the local barrier of the objective, effectively steering the CIM toward the feasible region while preserving maximum sensitivity to objective-function differences.

\section{Scalarization}
\label{sec:scalarization}

We first show that the weighted-sum scalarization commutes with the regression step, so the aggregated surrogate can be assembled directly from pre-computed individual coefficients without refitting.
Assume all $M$ objectives share the same polynomial basis. 
Since the ordinary least squares regression is a linear operation on the response data, the coefficients of the aggregated objective satisfy
\begin{equation}
\label{eq:beta}
\boldsymbol{\beta}^{(\boldsymbol{w})} = \sum_{i=1}^{M} w_i \boldsymbol{\beta}^{(i)},
\end{equation}
where $\boldsymbol{\beta}^{(i)}$ denotes the coefficient vector fitted independently for each objective.
Furthermore, because the binary encoding in Eq.~\eqref{eq:encoding} is a fixed linear mapping from $\bm{q}$ to $\bm{x}$, every entry of the resulting QUBO matrix is a linear function of $\bm{\beta}$. 
The superposition therefore propagates through the encoding and results in Eq.~\eqref{eq:QUBOM}. 

We next show that the Rosenberg penalty coefficients can be aggregated. 
Because all objectives share the same polynomial basis, the Rosenberg reduction introduces an identical set of auxiliary variables and constraints across objectives. 
For the aggregated objective $H^{(\boldsymbol{w})}(\boldsymbol{q}) = \sum_{i=1}^{M} w_i H^{(i)}(\boldsymbol{q})$, the required $r$-th penalty coefficient $\lambda_{\boldsymbol{w},r}^*$ can be bounded by applying the triangle inequality and the subadditivity of the maximum function:
\begin{equation}
\lambda_{\boldsymbol{w},r}^* 
= \eta \max_{k \in \{1,\dots,N\}} \left| \sum_{i=1}^{M} w_i \Delta_k H^{(i)}_{r}(\boldsymbol{q}) \right| 
\le \sum_{i=1}^{M} w_i \left( \eta \max_{k \in \{1,\dots,N\}} \left| \Delta_k H^{(i)}_{r}(\boldsymbol{q}) \right| \right) 
= \lambda^{(\boldsymbol{w})}_{r}.
\label{eq:lambdaw}
\end{equation}
Equation~\eqref{eq:lambdaw} guarantees that for every auxiliary constraint,  the weighted sum of the individually computed penalty coefficients $\lambda^{(\boldsymbol{w})}_{r}$ provides a safe upper bound. 
It ensures that no constraint is violated in the composite formulation of Eq.~\eqref{eq:QUBOM}.

\section{Hardware-aware precision adaptation}
\label{sec:split}
\setcounter{figure}{0}

The programmable coupling coefficients in the CIM are restricted to discrete values with finite bit-widths. 
Specifically, the CIM used in this study requires matrix coefficients to be mapped onto 8-bit signed integers~\cite{KaiwuSDK}. 
However, the QUBO matrices constructed from aerodynamic optimization inherently possess high-precision floating-point coefficients. 
Direct scaling and rounding distorts the energy landscape and can shift the location of the optimal solution. 
Therefore, a hardware-aware precision adaptation strategy is required. 

We adopt the PAS method, a topological expansion strategy that trades spin overhead for exact solution preservation. 
The floating-point QUBO matrix is first scaled by a factor $1/\epsilon$, where $\epsilon$ is the minimum precision threshold, to convert all entries into integers.
When a resulting coefficient $C$ exceeds the hardware range $R_{\max}$, the corresponding variable $q$ is replaced by $N_q$ redundant copies $\lrc{q^{(i)}}_{i=1}^{N_q}$. 
The weight is distributed evenly among the copies, and penalty terms enforce consistency
\begin{equation}\label{eq:split}
    Cq = \sum_{i=1}^{N_q} C_i q^{(i)} + \sum_{i < j}^{N_q} \mu_{ij} (q^{(i)} + q^{(j)} - 2q^{(i)}q^{(j)}),
\end{equation}
where $\sum_{m=1}^{N_q} C_m = C$ and the penalty factor $\mu_{ij}$ is chosen to be large enough to constrain all copies to the same value while satisfying $|-2\mu| \le R_{\max}$. 
The same splitting procedure applies to quadratic interaction terms between different variables whose weights exceed $R_{\max}$. 

\begin{figure}[!ht]
    \centering
    \includegraphics[width=\linewidth]{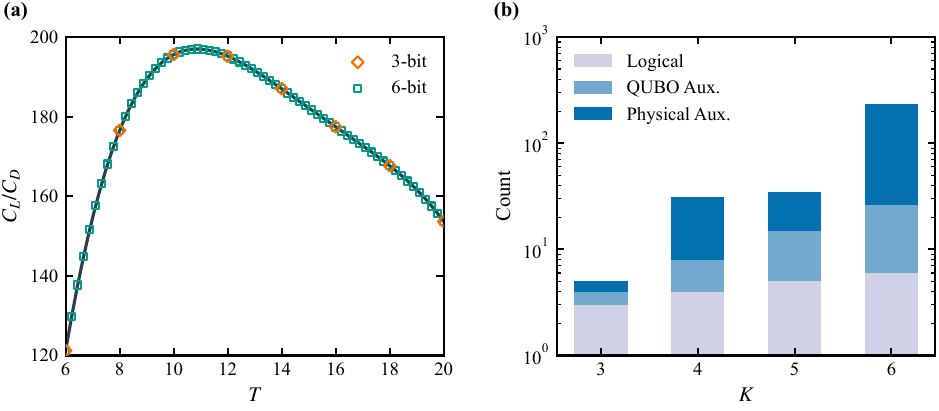}
    \caption{Analysis of solution candidates and spin overhead across different discretization precisions. (a) Comparison of discretization candidates under 3- and 6-bit precisions. (b) The spin counts for discretization bits $K=$ 3, 4, 5, and 6. 
    The total required spins are decomposed into the logical, QUBO auxiliary, and physical auxiliary spins.}
    \label{fig:precision}
\end{figure}

Figure~\ref{fig:precision} illustrates the spin overhead for the case study in Sec.~\ref{sec:case_HOBO} across discretization precisions of $K=$ 3, 4, 5, and 6 bits. 
As shown in Fig.~\ref{fig:precision}a, increasing the number of discretization bits from $K=$ 3 to 6 yields denser candidate solutions within the design space, enabling higher solution accuracy.
However, this improvement incurs a growing spin overhead, as presented in Fig.~\ref{fig:precision}b, where the total spin count rises from 5 to 233 across $K=$ 3 to 6. 
The total spins decompose into three categories: the logical spins encoding the design variables, the QUBO auxiliary spins introduced by the Rosenberg reduction, and physical auxiliary spins introduced by PAS. 
The PAS-induced overhead dominates at high precisions.
This overhead is specific to the current 8-bit hardware and is expected to diminish substantially with future improvements in the native coefficient resolution of CIM devices.

\bibliographystyle{elsarticle-num}
\biboptions{sort&compress}
\bibliography{QAirfoilOpt}

\end{document}